\documentstyle[12pt]{article}
\newlength{\pubnumber} 
\newcommand\pubblock[2]{\begin{flushright}\parbox{\pubnumber}
 {\begin{flushleft}#1\\ #2\\ \end{flushleft}}\end{flushright}}
\catcode`\@=11
\@addtoreset{equation}{section}
\def\section{\@startsection{section}{1}{\z@}{3.5ex plus 1ex minus .2ex}
 {2.3ex plus .2ex}{\large\bf}}
\def\subsection{\@startsection{subsection}{2}{\z@}{2.3ex plus .2ex}
 {2.3ex plus .2ex}{\bf}}

\begin{document}
\bibliographystyle{unsrt}
\begin{titlepage}
\samepage{
\setcounter{page}{1}
\pubblock{UFIFT-HET-95-31}{\tt hep-th/9512107}
\vfill
\begin{center}
 {\Large \bf Universality and Clustering in 1+1
Dimensional Superstring-Bit Models\footnote
{Supported in part by
the Department of Energy under grant DE-FG05-86ER-40272,
and by the Institute for Fundamental Theory.}}
\vfill
 {\large Oren Bergman\footnote{E-mail  address: oren@phys.ufl.edu} and
             Charles B. Thorn\footnote{E-mail  address: thorn@phys.ufl.edu}
   \\}
\vspace{.12in}
 {\it  Institute for Fundamental Theory\\
Department of Physics, University of Florida, Gainesville,
FL, 32611, USA}
\end{center}
\vfill
\begin{abstract}
\noindent We construct a 1+1 dimensional superstring-bit
model for D=3 Type IIB superstring. This low dimension
model escapes the problems encountered in higher
dimension models: (1) It possesses full Galilean
supersymmetry; (2) For noninteracting
polymers of bits, the exactly soluble linear
superpotential describing
bit interactions
is in a large universality class of
superpotentials which includes ones bounded at spatial infinity;
(3) The latter are used to construct a superstring-bit
model with the clustering properties needed to define
an $S$-matrix for closed polymers of superstring-bits.
\end{abstract}
\vfill
\smallskip}
\end{titlepage}

% ========================= DEFINITIONS ===================================
\def\beq{\begin{equation}}
\def\eeq{\end{equation}}
\def\beqn{\begin{eqnarray}}
\def\eeqn{\end{eqnarray}}
%==============================================================================
\hyphenation{su-per-sym-met-ric non-su-per-sym-met-ric}
\hyphenation{space-time-super-sym-met-ric}
\hyphenation{mod-u-lar mod-u-lar--in-var-i-ant}
%==============================================================================

%============================== TEXT BEGINS HERE ============================

\setcounter{footnote}{0}
\bibliographystyle{unsrt}
%=========================================================================
%=========================================================================
\def\balpha{\mbox{\boldmath$\alpha$}}
\def\bgamma{\hbox{\twelvembf\char\number 13}}
\def\bsigma{\hbox{\twelvembf\char\number 27}}
\def\bepsilon{\hbox{\twelvembf\char\number 15}}
\def\sgone{{\cal S}_1{\cal G}}
\def\sgtwo{{\cal S}_2{\cal G}}
\def\ob{\overline}
\def\bx{{\bf x}}
\def\by{{\bf y}}
\def\bz{{\bf z}}
\def\D{{\cal D}}
\def\R{{\cal R}}
\def\Q{{\cal Q}}
\def\G{{\cal G}}
\def\O{{\cal O}}
\def\Nlarge{N_c\rightarrow\infty}
\def\Tr{{\rm Tr}}
\newcommand{\ket}[1]{|#1\rangle}
\newcommand{\bra}[1]{\langle#1|}
\newcommand{\firstket}[1]{|#1)}
\newcommand{\firstbra}[1]{(#1|}

String-bit models are
attempts to reformulate string theories
in $D-1$ space dimensions
as
field theories of point-like string-bits moving in $d\equiv D-2$
space dimensions,
in which string is composite, not fundamental
\cite{gilest,klebanovs,thornmosc,thornrpa,bergmantbits}.
The bits bind together
to form long closed polymer chains,
which in the continuum limit have precisely the properties of
closed relativistic string. String-bits
must carry an internal ``color'' degree
of freedom which defines ordering around the chain. This can be
achieved
if the bits transform in the adjoint
representation of the unitary group $U(N_c)$, with $N_c\geq 2$.
A crucial feature of our light-cone approach
to string-bit models is that they possess
Galilean invariance in $d$ space dimensions,
not
Poincar\'e invariance in $d+1$ space dimensions: the
bits enjoy a {\bf non-relativistic} dynamics.

Second-quantization of string-bits
employs a string-bit
creation operator $\phi^\dagger(x)^\beta_\alpha$, where
$\alpha$ and $\beta$ run over the $N_c$ colors and $x$ denotes the
$d$ space coordinates.
Denote the zero-bit state by $\ket{0}$.
A bare closed chain of $M$ bits is then described by:
\begin{eqnarray}
 \ket{\Psi(&x_1&,\ldots,x_M)} =\int dx_1\cdots dx_M \nonumber\\
 &&~~~~~~\times\Tr[\phi^\dagger(x_1)\cdots\phi^\dagger(x_M)]\ket{0}
     \Psi(x_1,\ldots ,x_M)\; ,
 \label{chain}
\end{eqnarray}
implying that $\Psi(x_1,\ldots ,x_M)$
is cyclically symmetric.
To describe closed chains and their interactions,
the Hamiltonian governing string-bit dynamics must allow
for their formation and assure their stability.

Chain formation requires that string-bits have an
attractive interaction
between nearest neighbors on a chain, with non-nearest neighbors
interacting much more weakly.
It is well-known \cite{thornfock,thornmosc,bergmantbits}
that this  pattern
of interactions arises in a many body system of
particles described by
$N_c\times N_c$ matrix creation operators using
't Hooft's $N_c\rightarrow\infty$ limit \cite{thooftlargen}.
For an interaction Hamiltonian of the form
\begin{equation}
 H_{int} = {1\over N_c}\int dxdy V(y-x)
   \Tr[\phi^\dagger(x)\phi^\dagger(y)\phi(y)\phi(x)],
\label{intham}
\end{equation}
the limit $\Nlarge$ leads to nearest-neighbor interactions in a
bare closed chain
of string-bits. For $N_c$ finite but large,
$O(1/N_c)$ effects
allow a single bare closed chain to break into two bare closed chains.
Thus $1/N_c$ serves as a
chain coupling constant, and 't Hooft's $1/N_c$ expansion
produces chain perturbation theory.

Free light-cone string is recovered for $\Nlarge$ in the continuum
limit  given by $M\rightarrow\infty, m\rightarrow 0$, with
$mM$ kept fixed, or equivalently in the low energy limit given
by $E\ll T_0/m$, where $T_0$ is the string tension and $m$ is
the Newtonian mass of a bit.
The total Newtonian mass of a chain becomes an
effectively continuous $P^+$
of a string. The $x^-$ coordinate of string thus
emerges dynamically in string-bit models as the
conjugate to Newtonian mass.
The other light-cone coordinate $x^+$ is
identified as time, and its conjugate $P^-$ as the bit
Hamiltonian.
The $O(1/N_c)$ chain interactions become string
interactions in the continuum limit.

Stability of string depends on the ground state energy of
a long closed chain, generically given by:
\begin{equation}
 E_{0,M} = {1\over m}\left[aM + {b\over M}
       + O({1\over M^2})\right].
\label{gsenergy}
\end{equation}
The first term is the same for a single chain of $M$ bits
and two chains of $M_1$ and $M_2$ bits, with $M_1+M_2=M$.
For long chains, the nature of the true ground state then depends on
the second term. If $b>0$ then
$E_{0,M}<E_{0,M_1}+E_{0,M_2}$,
and a long chain is stable. If $b<0$ then
$E_{0,M}>E_{0,M_1}+E_{0,M_2}$, the chain is unstable to
decay into two smaller chains, and
it will through the $O(1/N_c)$
terms alluded to before.
Consider
a chain of bosonic bits interacting via the
nearest-neighbor harmonic
potential $V(x)=(\omega^2/2m)x^2$.
The ground state energy for an $M$-bit
chain in
$d$ space dimensions is found to be
\cite{gilest,bergmantbits}
\begin{equation}
 E_{0,M} = {\omega d\over 2m}\sum_{n=1}^{M-1}\sin{n\pi\over M}
            = {\pi\omega d\over m}\left[{2}M
              - {1\over 6M} + O({1\over M^2})\right],
\label{harmonicgs}
\end{equation}
and so a long
chain of bosonic bits is unstable against decay into
two smaller chains. This is just the string-bit manifestation
of the tachyonic instability of bosonic string.
The negative coefficient of $1/2mM$ is the mass-squared
of the tachyon.

This instability is absent in  superstring theory
which requires the addition of fermionic modes on string
in a supersymmetric fashion. For string-bit models
the bits are in supermultiplets with a
``statistics'' degree of
freedom distinguishing bosons from fermions. This
degree of freedom gives rise to ``statistics waves'' on
long chains,
similar to spin waves.
Supersymmetrizing
the harmonic string-bit
model leads to a cancellation of the contribution
to the ground state energy
of the coordinate ``phonon'' waves
with that of the ``statistics''
waves.
In fact,
the ground state energy
is exactly zero
for any $M$\cite{bergmantbits}.
We shall see later that
this is a universal property of supersymmetric
string-bit models, and not special to the harmonic interaction.
Note that recovering the free superstring mass spectrum at
$M=\infty$ requires only $b=0$, whereas in
this model
{\bf all} the terms in (\ref{gsenergy}) vanish. If this
were not so, the sign of the first nonvanishing term would
determine the stability of finite long chains. If these were
unstable except when $M=\infty$, the
bit model could not provide a {\bf fundamental}
basis for superstring theory, and
at best
would only make sense in the continuum limit.

We have constructed superstring-bit models in $2+1$
and $8+1$ dimensions that underlie $D=4$ and $D=10$ type IIB
superstring theory respectively \cite{bergmantbits}. In the $D=4$
case we
could include
extra degrees of freedom
either as real compactified dimensions, or
as additional internal bit degrees of freedom \cite{gilesmt} which,
on long chains, would produce ``flavor waves'' playing the role of
extra dimensions.
The $N=2$ spacetime Poincar\'e supersymmetry of the $D$-dimensional
type IIB superstring requires the corresponding $(D-2)+1$-dimensional
superstring-bit model to possess an $N=1$ Galilean supersymmetry.
The symmetry is Galilean because light-cone
variables
break the
manifest $SO(D-1,1)$
to $SO(D-2)\times SO(1,1)$.
Discretization of $P^+$ breaks this
$SO(1,1)$ and also mixes the right-
and left-moving supercharges,
leaving only an $N=1$ supersymmetry generated by the
right $+$ left combinations.
The Galilean supercharges $\Q$ and $\R$ transform as spinors
of the transverse $SO(D-2)$ subgroup of $SO(D-1,1)$,
and have opposite chirality in the $SO(1,1)$ subgroup.
Together they build a single supercharge transforming
as a spinor of the Lorentz group $SO(D-1,1)$, and generating
the super-Poincar\'e algebra.

For general $d$ the Galilean supercharges $\Q^A,\R^{\dot A}$ must
each have $d$ components for a satisfactory superstring limit of
the string-bit model. The corresponding super-Galilei
algebra then reads:
\begin{eqnarray}
 \{\Q^A,\Q^B\} = mM\delta^{AB}\; &,& \;
 \{\Q^A,\R^{\dot B}\} = {1\over 2}\balpha^{A\dot B}\cdot {\bf P}\; ,
    \nonumber\\
  \{\R^{\dot A},\R^{\dot B}\} &=& \delta^{{\dot A}{\dot B}}H/2 \; ,
\label{twoplusonealg}
\end{eqnarray}
where $M$ is the total number of bits,
and ${\bf P}$ is their total momentum.
The superstring-bit models of \cite{bergmantbits} implement
all but the last of these relations:
There are additional
terms not proportional to $\delta^{{\dot A}{\dot B}}$.
For the supersymmetric harmonic model these terms are
sub-leading in the $1/N_c$ expansion, so the $\R$-supersymmetry
is broken by chain interactions. For other superstring-bit models
this happens already at the level of the free chain spectrum.
In either case, it might still be that the full Poincar\'e supersymmetry
can be recovered in the continuum stringy physics.

For $d=1$, the full
superalgebra closes by default,
since $\R$ and $\Q$ then have only one component each.
A $1+1$-dimensional
superstring-bit model would underlie $D=3$ superstring,
the lowest dimensional superstring possible \cite{greensw}.
Specializing the supercharges
of \cite{bergmantbits} to this case gives
\begin{eqnarray}
 \Q &=& \sqrt{m\over 2}\int dx
  \Tr\big[e^{i\pi/4}\phi^\dagger(x)\psi(x) + {\rm h.c.}\big]\nonumber\\
 \R &=& -{1\over 2\sqrt{2m}}\int dx
  \Tr\big[e^{-i\pi/4}\phi^\dagger(x)\psi^\prime(x) + {\rm h.c.}\big]\nonumber\\
 & & + {1\over2N_c\sqrt{2m}}\int dx dy W(y-x) \nonumber\\
 & &~~~~~~~~~~\times\Tr\big[e^{-i\pi/4}\phi^\dagger(x)\rho(y)\psi(x)
     + {\rm h.c.}\big]\; ,
\label{secondquant}
\end{eqnarray}
where
$\phi^\dagger(x)_\alpha^\beta$ is the bosonic creation operator,
$\psi^\dagger(x)_\alpha^\beta$ is the fermionic creation operator, and
$\rho^\beta_\alpha = [\phi^\dagger\phi +\psi^\dagger\psi]^\beta_\alpha$.
The superalgebra is given by:
\begin{eqnarray}
  \{\Q,\Q\} = mM&,& \;
  \{\Q,\R\} = -P/2, \;
  \{\R,\R\} = H/2.
\label{anticomm}
\end{eqnarray}
The last equation can be taken as the {\bf definition}
of the Hamiltonian
for this system, which
in turn implies that the model is invariant
under both $\Q$ and $\R$.

If the function $W(x)$ is taken to be odd, the two-bit
sector is equivalent to Witten's supersymmetric
quantum mechanics \cite{wittensusyqm}, where $W(x)$
is the superpotential. In that case the ground state
energy of a two-bit closed chain vanishes.
For understanding superstring theory,
we are interested in a class of superpotentials
for which the ground state energy of {\bf any}
length chain vanishes, and for which the gap to
excite the chain is finite.
Exploring this issue for noninteracting chains,
we consider
the first-quantized system obtained by acting
with $\R$ on a bare chain and taking the limit $\Nlarge$.
The first-quantized supercharge is then given by
\begin{eqnarray}
 R &=& {1\over 2\sqrt{2m}}
       \sum_{k=1}^M\Big\{
          (e^{i\pi/4}\theta_k+e^{-i\pi/4}\pi_k)p_k \nonumber \\
   & &~~~~~~~-(e^{-i\pi/4}\theta_k+e^{i\pi/4}\pi_k)W(x_{k+1}-x_k)
   \Big\}\; ,
\label{onedsusy}
\end{eqnarray}
where $p_k=-i\partial/\partial x_k$ and
$\pi_k = \partial/\partial\theta_k$. The summation is understood
to be cyclic, i.e. $k=M+1$ is equivalent to $k=1$.
The first-quantized M-bit Hamiltonian is then given by
\begin{eqnarray}
 H_M &=& {1\over 2m}\sum_{k=1}^M\Big\{
   p_k^2 + W^2(x_{k+1}-x_k) \nonumber\\
 &+& W^\prime(x_{k+1}-x_k)[\theta_k\pi_{k}
                                 - \pi_k\theta_{k}\nonumber\\
 &+& \pi_{k+1}\theta_k - \theta_{k+1}\pi_k-i(\theta_k\theta_{k+1}
+\pi_k\pi_{k+1})] \Big\}.
\label{firstham}
\end{eqnarray}
We denote states in the first-quantized Hilbert space
of an $M$-bit chain by $\firstket{\cdots}$ to distinguish
them from states in the second-quantized bit Fock space,
denoted $\ket{\cdots}$. The ground state of the chain
is then denoted by $\firstket{0}$.

Consider the linear superpotential
\begin{equation}
 W(x)=T_0x \; ,
\label{linear}
\end{equation}
which defines the supersymmetric harmonic model.
The ground state is annihilated by $R$ implying it belongs
to a ``small'' representation of the Galilei
superalgebra and has zero energy.
In addition its spectrum approaches
that of a relativistic superstring with tension $T_0$ in the
continuum limit.
Consider a deformation of the superpotential
\begin{equation}
 W(x)\rightarrow W(x) + \delta W(x) \; ,
\label{deform}
\end{equation}
where $\delta W(x)$ is small in the interval $|x|<L$.
If $L\gg 1/\sqrt{T_0}$ this deformation can be treated
perturbatively,
since for $|x|>L$ the unperturbed
wavefunctions are exponentially small.
Due to the Galilei superalgebra (\ref{anticomm}),
the exact Hamiltonian can be written as the square
of the new supercharge $R + \delta R$.
The change in the energy
of the ground state to first order is then:
\begin{equation}
 \delta E_{0,M} = 4\firstbra{0}\{R,\delta R\}\firstket{0} \; ,
\label{gsshift}
\end{equation}
which vanishes since $R\firstket{0}=\firstbra{0}R=0$.
We stress that this holds for {\bf any} length
chain.

The spectrum of excitations of the chain is generated
by acting on the above ground state with mode
raising operators.
For the $2+1$-dimensional harmonic model these were derived
in \cite{bergmantbits}. Dropping a dimension and with it the
spinor
indices gives:
\begin{equation}
   A^\dagger_n = {({\hat p}_n
              + i\omega_n{\hat x}_n)\over\sqrt{2\omega_n}} \; , \;
   A_n = {({\hat p}_n
              - i\omega_n{\hat x}_n)\over\sqrt{2\omega_n}} \; ,
\end{equation}
where $\omega_n = 2T_0\sin n\pi/M$, for the coordinate modes
raising and lowering operators, and
\begin{equation}
  B^\dagger_n = \xi_n{\hat\theta}_n + \eta_n{\hat\pi}_n \; , \;
  B_n = \eta_n{\hat\theta}_n + \xi_n{\hat\pi}_n \; ,
\end{equation}
where $\xi_n = (1/\sqrt{2})(\sin{n\pi/2M} + \cos{n\pi/2M})$ and
$\eta_n = (1/\sqrt{2})(\sin{n\pi/2M} - \cos{n\pi/2M})$,
for the ``statistics'' modes
raising and lowering
operators.
In the above ${\hat x}_n,{\hat\theta}_n,{\hat p}_n,{\hat\pi}_n$
are the Fourier transforms of $x_k,\theta_k,p_k,\pi_k$,
respectively.
Consider an excitation with a single raising operator
$A^\dagger_n\firstket{0}$.
The zero'th order energy of this state is given by
\begin{equation}
 E^{(0)}_{n,M} = \firstbra{0}A_nHA^\dagger_n\firstket{0}
               = {2T_0\over m}\sin{n\pi\over M}\; .
\label{energynM}
\end{equation}
The shift in the energy due to the deformation
(\ref{deform}) is given to first order by
\begin{equation}
 \delta E_{n,M} =
    4\firstbra{0}A_n\{R,\delta R\}A^\dagger_n\firstket{0}\ .
\end{equation}
Using the commutation relations
\begin{eqnarray}
 [A_n,R]&=& -{i\over 2}e^{i\pi n/ 2M}
      \sqrt{\omega_n\over m}B_n\nonumber\cr
 [A_n,\delta R]&=&
  {-\sin n\pi/M\over 2\sqrt{\omega_nmM}}\sum_k\Big\{
     (e^{-i\pi/4}\theta_k + e^{i\pi/4}\pi_k)
     \\
   &&~~~~~~~~\times  e^{i\pi n(2k+1)/M}
      \delta W^\prime(x_{k+1}-x_k)\Big\}\; ,
\end{eqnarray}
and the fact that $\firstbra{0}\delta W^\prime(x_{k+1}-x_k)\firstket{0}$
is cyclically invariant
we find
\begin{eqnarray}
 \delta E_{n,M} &=& -2ie^{i\pi n/2M}\sqrt{\omega_n\over m}
    \firstbra{0}B_n\delta RA^\dagger_n\firstket{0}
 + {\rm h.c.}\nonumber\\
 &=& {2\over m}\firstbra{0}\delta W^\prime(x_2-x_1)\firstket{0}
        \sin{n\pi\over M}\; .
\label{nshift}
\end{eqnarray}
The net effect is then just to shift the string tension
\begin{equation}
 T_0\rightarrow T_0
   + \firstbra{0}\delta W^\prime(x_2-x_1)\firstket{0} \; ,
\label{newtension}
\end{equation}
so the gap remains finite for all $M$.
For excitations of the form $\prod_i A^\dagger_{n_i}\firstket{0}$,
with more than one raising operator,
we have
\begin{eqnarray}
 \delta E &=& 4\firstbra{0}\prod_i A_{n_i}\{R,\delta R\}
             \prod_i A^\dagger_{n_i}
             \firstket{0}\nonumber\\
 &=& -2i\sum_je^{i\pi n_j/2M}\sqrt{\omega_{n_j}\over m}
      \firstbra{0}B_{n_j}\prod_{i\neq j}A_{n_i}\delta R
      \prod_k A^\dagger_{n_k}\firstket{0} \nonumber \\
 &&+{\rm h.c.}\; .
\end{eqnarray}
As we commute the $A^\dagger_{n_k}$'s to the left we
pick up an additional derivative of $\delta W$ and
a factor of $\sqrt{\omega_{n_k}/M}=O(1/M)$
for each $A^\dagger_{n_k}$ that contracts with $\delta R$.
The rest contract with some of the $A_{n_i}$'s. The remaining
$A_{n_i}$'s are then commuted to the right, all
contracting with $\delta R$ to produce additional
derivatives and additional powers of
$1/{M}$.
Finally, $B_{n_j}$ is contracted with what's left.
The result is a sum of terms of increasing odd number of
derivatives of $\delta W$
multiplied by increasing powers of $1/M$, the coefficient
of $\delta W^{(2l+1)}$ being proportional to $M^{-2l-1}$.
In the limit $M\rightarrow\infty$ only the
the $l=0$ term contributes, so the analogue of (\ref{nshift})
holds for any excitation of the form
$\prod_i A^\dagger_{n_i}\firstket{0}$.
The argument for excitations created by products of
$B^\dagger_n$'s or products of both $A^\dagger_n$'s
and $B^\dagger_n$'s is
similar. Thus
the only effect of the deformation (\ref{deform}) on long
chains is to
renormalize the string tension as in Eq.(\ref{newtension}).

We now argue that these results from first order perturbation
theory hold to all orders, and indeed should extend to a large
universality class of superpotentials. The ground state energy
must remain zero as long as the ground state is in a ``small''
representation of the superalgebra. Clearly the representation
can't change in perturbation theory, but more generally it
will remain ``small'' unless the first excited
state becomes degenerate with the ground state, i.e. unless
the gap closes. Moreover, the properties of the $O(T_0/mM)$
excitations of long chains have a universal character
determined by phonon and statistics waves, which
are inevitable collective excitations of stable long chains
\cite{thornrpa}.
Formation of long bare chains is
ensured by a
large bond-breaking energy and does not require
the infinite range harmonic force.

Finally we turn to the issue of interactions between
closed chains. By exploiting the $N_c\to\infty$ limit,
we have been able to define a {\bf bare} closed chain, whose interactions
with other closed chains is negligible.
Taking $N_c$ to be finite will
``dress''
the bare chains, and will give rise to interactions
between different chains.
The ionization
energy required to break a bond in the closed chain
is $O(T_0/m)$. As long as the scattering energy is below
threshold for such ionization, the only way for interactions
to occur is through bond rearrangement.
A necessary condition for defining a closed chain
$S$-matrix is that the model satisfy a {\bf clustering}
property:
An initial state
of two spatially separated closed chains
must evolve as
two noninteracting chains
until enough time has elapsed for them to get close to
one another.
With a linearly growing superpotential as in
(\ref{linear}), this is impossible at finite $N_c$.
The supersymmetric harmonic model is thus unsatisfactory
for describing chain scattering.
Asymptotically free chains require a bounded
superpotential.
In order to keep the ionization energy $O(T_0/m)$, however,
we must
still insist that
$W\to\pm W_{\infty}\neq0$
at spatial infinity.
The restricted universality
established above indeed allows us to deform the linear
superpotential into
\begin{eqnarray}
W=T_0\big[x+(L-x)\theta(x-L)-(L+x)\theta(-L-x)\big],
\label{bounded}
\end{eqnarray}
with no change in the continuum properties of free chains.

With the bounded superpotential (\ref{bounded}) there can still
be large ($O(T_0/m)$) correlation
energies between spatially separated clumps of bits.
If the clumps correspond to closed chains, or
more generally to singlet states, these correlation
energies must vanish. This can be achieved by first
noting that
the generators for color rotations given by
\begin{eqnarray}
G_\alpha^\beta=\int dx \big[\rho(x)-
\sigma(x)\big]_\alpha^\beta\; ,
\end{eqnarray}
where
$\sigma_\alpha^\beta=
:[\phi\phi^\dagger-\psi\psi^\dagger]_\alpha^\beta:$,
annihilate singlet states.
This motivates
replacing $\rho(y)$ with $\rho(y)-\sigma(y)$ in Eq.
(\ref{secondquant}) for $\R$.
It is easy to check that such a change does not disturb
the superalgebra.
To see that it gives the desired clustering
property, consider the action of the new $\R$ on a state we
denote by
$S_1S_2\ket{0}$,
where $S_1$ and $S_2$ are any color singlet
functions of creation
operators, such that the locations of all creation operators in $S_1$
are more than a distance $L$ from all the locations in $S_2$.
Let $(rS)_1$ and $(rS)_2$ be the singlet
functions of creation
operators defined by
\begin{equation}
\R S_1\ket{0}=(rS)_1\ket{0}\; , \;
\R S_2\ket{0}=(rS)_2\ket{0}\ .
\end{equation}
Then
\begin{equation}
\R S_1S_2\ket{0}=(rS)_1S_2\ket{0}+(-)^{S_1}S_1(rS)_2\ket{0}
 +\R_IS_1S_2\ket{0}
\end{equation}
where the action of $\R_I$ is defined by requiring
one of the two annihilation operators in the two-body term of $\R$
to contract
with a creation operator in $S_1$, and the other annihilation operator
to contract with a creation operator in $S_2$.
Because of the spatial separation of the coordinates in $S_1$
and the coordinates in $S_2$, the superpotential $W(x)$ can
be replaced with its asymptotic value and taken out of the
integral. Consequently one is left with the color rotation
operator $\int(\rho-\sigma)$ acting on $S_1$ or $S_2$, either
of which gives zero.
Thus we conclude that
\begin{equation}
\R S_1S_2\ket{0}=(rS)_1S_2\ket{0}+(-)^{S_1}S_1(rS)_2\ket{0}\ .
\end{equation}
Applying $\R$ once again, using the supersymmetry algebra, and
remembering that the supercharge is odd, we infer:
\begin{equation}
 HS_1S_2\ket{0}=4\bigl[(r^2S)_1S_2\ket{0}+S_1(r^2S)_2\ket{0}\bigr],
\end{equation}
implying that the hamiltonian acts independently
on the two singlets.
Therefore as long as they remain spatially well separated the two singlets
propagate without mutual interaction.

%This work was supported in part by
%the Department of Energy under grant DE-FG05-86ER-40272,
%and by the Institute for Fundamental Theory.

\end{document}